\documentclass[twocolumn]{revtex4}
\usepackage[dvipdfmx]{graphicx}% Include figure files
\usepackage{amsmath,amsbsy,amssymb}
\usepackage{bm}% bold math
\usepackage{mathrsfs}
\usepackage{ulem}
\usepackage{textgreek}
\usepackage{multirow}

\usepackage{color}

\newcommand{\diff}{\mathrm{d}}

\newcommand{\trace}{\mathrm{Tr}\,}
\newcommand{\imu}{\mathrm{i}}
\newcommand{\epn}{\mathrm{e}}

\newcommand{\dg}{\dagger}
\newcommand{\la}{\langle}
\newcommand{\ra}{\rangle}
\newcommand{\al}{\alpha}

\newcommand{\gm}{\gamma}
\newcommand{\ep}{\varepsilon}
\newcommand{\percent}{\%}

\begin{document}

\title{
Electronic 
multipoles and multiplet pairs
induced 
from \\
Pomeranchuk and Cooper instabilities of
Bogoliubov Fermi surfaces
}

\author{Shun-Ta Tamura, Shoma Iimura and Shintaro Hoshino}

\affiliation{
Department of Physics, Saitama University, Shimo-Okubo, Saitama 338-8570, Japan
}

\date{\today}

\begin{abstract}
It has recently been pointed out that Fermi surfaces can remain even in the superconductors under the symmetric spin-orbit interaction and broken time-reversal symmetry.
Using the linear response theory, we study the instability of such systems toward ordering, which is an intrinsic property of the Fermi surfaces.
The ordered states are classified into diagonal and offdiagonal ones, each of which respectively indicates the Pomeranchuk instability and Cooper pairing not of original electron but of Bogoliubov particles (bogolons).
The corresponding order parameters are expanded by multipole moments (diagonal order parameter) and multiplet pair amplitudes (offdiagonal order parameter) of original electrons, which are induced by the internal fields arising from bogolons' ordering.
While the 
bogolons' order parameters
partially inherit the characters of the original electrons, many order parameter components mix with similar magnitude.
Hence 
there is no clear-cut distinction whether the phase transition is diagonal or offdiagonal ordering in terms of the original electrons.
These ordering instabilities inside the superconducting states provide insights into the superconductors which have the second phase transition below the first transition temperature.
\end{abstract}

\maketitle

\section{Introduction}

The Fermi surfaces are known to show a variety of intriguing phenomena at low temperatures because of the interaction effects among electrons.
A typical example is the magnetic ordering, where the spin degrees of freedom near the Fermi energy are reconstructed by repulsive interactions.
The transition temperature where the order parameter, or magnetization, becomes finite is proportional to the density of states known as the Stoner criterion. 
It has also been recognized that the presence of the Fermi surfaces drives the system to the nonmagnetic ordering, which results in spatial-symmetry lowering (Pomeranchuk instability \cite{Pomeranchuk58}).
The other examples where the Fermi surface effect is involved in mechanisms include the screening of the magnetic moments in metals known as the Kondo effect \cite{Kondo64,Yosida66}.

The superconducting state, which is caused by the Cooper pair formation \cite{Cooper56,Bardeen57}, is also a manifestation of the instability of the Fermi surfaces, whose elementary fermionic excitations are described as the emerging Bogoliubov quasiparticles (or bogolons) written in terms of the electron-hole superposed state \cite{Bogoliubov58}.
The instability toward superconductivity is a
quite general phenomenon as evidenced by the logarithmically divergent pairing susceptibility at low temperatures even without interactions due to the presence of Fermi surfaces.
Usually, the Fermi surfaces have gone in the resultant pairing state, and 
the system reaches the ground state with no more degrees of freedom left for electrons.

On the other hand, an interesting possibility has been pointed out theoretically:
the Bogoliubov particles can form stable Fermi surfaces in some superconductors (Bogoliubov Fermi surfaces) \cite{Volovik89,Liu03,Gubankova05,Agterberg17,Brydon18,Yuan18}.
As discussed above, the remaining Fermi surface has instabilities toward ordering, and here it appears {\it inside the superconducting state}.
Because of the intrinsic logarithmically divergent pair susceptibility, it is naturally expected that the system with Bogoliubov Fermi surfaces shows a pairing state of bogolons at sufficiently low temperatures.
Such possibility is indeed studied recently \cite{Oh19}.
In the present paper, we study the properties of both diagonal (Pomeranchuk instability) and offdiagonal (Cooper instability) order parameters of bogolons.
Specifically for the Cooper pairing of boglons, since the pairing state of original electrons is already realized, the nature of the second pairing state inside the superconducting state are interesting but unclear.
We discuss this ``pairing state of pairing state'' based on a simple model that shows Bogoliubov Fermi surfaces.
These insights can also provide a candidate scenario for the ordered states inside the superconducting state.

In this paper, using the $j=3/2$ electron model \cite{Brydon16,Agterberg17,Menke19,Kobayashi19} with Bogoliubov Fermi surfaces, we identify emergent order parameters of the original electrons that are induced from bogolons' diagonal and offdiagonal orderings.
For this purpose, we employ the multipole expansion in terms of original electron degrees of freedom. 
The 
concept of
electronic multipoles originates from the spin-orbital model \cite{Kugel72} in $d$-electrons systems, and has been further extended to $f$-electron and other systems \cite{Ohkawa83, Shiina97, Kuramoto00, Santini00, Takimoto05, Kusunose08, Kuramoto09, Suzuki17, Hayami18}.
Since the pairing amplitudes are also involved in the superconducting state, we need to extend the multipole expansion to include pair amplitudes on equal footing.
As a result, the bogolon order parameters are represented by the mixture of diagonal and offdiagonal order parameters of original electrons.
Hence, while the ordering of bogolons is a good physical picture, 
we
cannot 
simply 
classify them as either diagonal or offdiagonal orders of the original electrons.

This paper is organized as follows.
In the next section we show the formulations for Bogoliubov Fermi surface and the order parameters.
The numerical results are given in Sec.~III.
We summarize the paper in Sec.~IV and make a comment on relevance to real materials.
The full lists of multipole operators, multiplet pair amplitudes and form factors in the wavevector space are given in Appendices A and B.

\section{Formulation}

\subsection{Model Hamiltonian}

We take the simplest model that has stable Bogoliubov Fermi surfaces as introduced in Ref.~\cite{Agterberg17}.
We consider the continuum model by focusing on a part of the Brillouin zone.
The Hamiltonian reads
\begin{align}
\mathscr H &= \sum_{\bm k} \vec c_{\bm k}^\dg \big[ \al (\bm k^2 - k_{\rm F}^2) \hat 1 + \beta (\bm k\cdot \hat{\bm J})^2 \big] \vec c_{\bm k}
\nonumber\\
&\ \ \ 
+ \sum_{\bm k} \Big(
\vec c_{\bm k}^\dg \hat \Delta_{\bm k} \vec c^{\dg{\rm T}}_{-\bm k} + {\rm H.c.}
\Big)
, \label{eq:ham_orig}
\end{align}
where $\vec c_{\bm k} = (c_{\bm k,\frac 3 2}, c_{\bm k,\frac 1 2}, c_{\bm k,-\frac 1 2}, c_{\bm k,-\frac 3 2})^{\rm T}$ is the spin-3/2 spinor composed of electron annihilation operators.
The hat ($\hat\ $) symbol represents a $4\times 4$ matrix.
$\hat{\bm J}$ is the angular momentum matrix for $j=3/2$ [see Eqs.~(\ref{eq:defJy}--\ref{eq:defJx})].
The parameter $\al$ is a constant proportional to the inverse of mass, $k_{\rm F}$ is the Fermi wavevector, and $\beta$ is the symmetric spin-orbit coupling.
The Fermi energy is given by $\ep_{\rm F} = \al k_{\rm F}^2$.
The gap parameter is chosen as 
\begin{align}
\hat \Delta_{\bm k} = 
  \Delta_1 
k_z (k_x + \imu k_y)
\hat E
+ \tfrac{2}{\sqrt{3}} \, \Delta_0 \big\lceil
\hat J_z (\hat J_x + \imu \hat J_y ) \big\rfloor
\hat E
,
\label{eq:def_Delta}
\end{align}
where the bracket $\lceil\cdots\rfloor$ symmetrizes the expression as $\lceil AB \rfloor = (AB+BA)/2$.
We have defined the antisymmetric tensor $\hat E$ [see Eq.~\eqref{eq:anti_sym_tensor}].
The above Hamiltonian guarantees the presence of the stable Fermi surfaces even in the superconducting state with $\Delta_{0,1}\neq 0$ \cite{Agterberg17}.
Hence we naively expect another phase transition at low enough temperatures.
Once the Hamiltonian is fixed in this way, we can move to a diagonalized picture as
\begin{align}
\mathscr H = \sum_{\bm k \in {\rm HBZ}}\vec \psi_{\bm k}^\dg \check H_{\bm k} \vec \psi_{\bm k}
= \sum_{\bm k \in {\rm HBZ}}\vec \al_{\bm k}^\dg \check \Lambda_{\bm k} \vec \al_{\bm k}
,
\end{align}
where $\vec\psi_{\bm k} = (\vec c_{\bm k}^{\rm T}, \vec c_{-\bm k}^\dg)^{\rm T}$ 
is the eight-component Nambu spinor, and $\vec \al_{\bm k} = \check U^\dg_{\bm k}\vec \psi_{\bm k}$ is Bogoliubov particle annihilation operators with a diagonal eigenvalue matrix $\check \Lambda_{\bm k}$ and eigenvector matrix $\check U_{\bm k}$.
The check ($\check \ $) symbol represents a $8\times 8$ matrix.
Note that the wavevector summation is taken over the half Brillouin zone (HBZ) to avoid double counting.

\begin{figure}[t]
\begin{center}
\includegraphics[width=85mm]{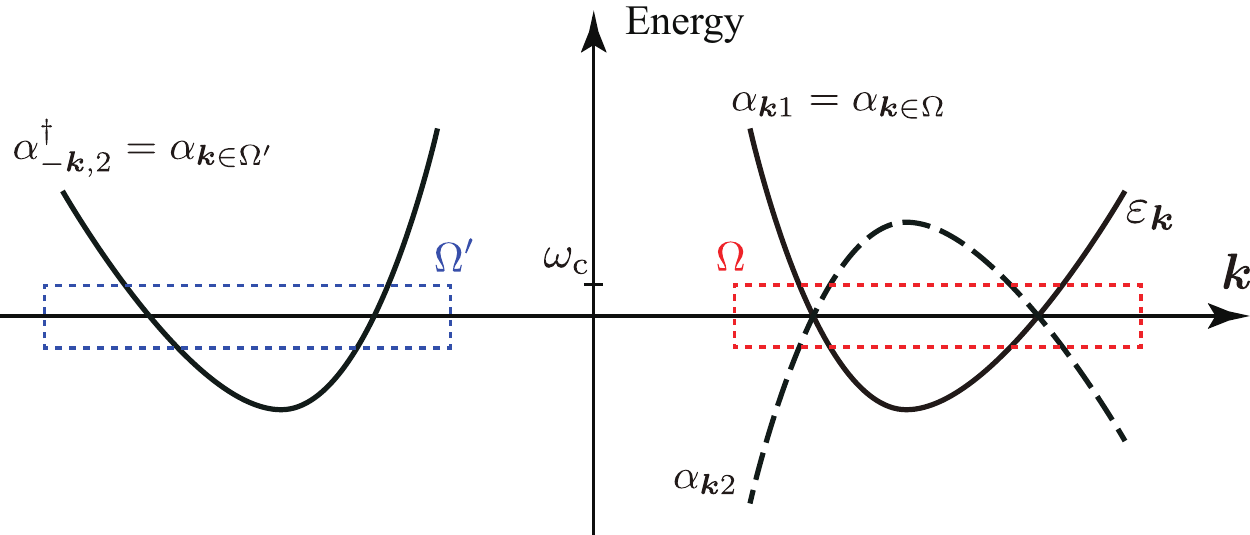}
\caption{
Illustration for the band structures of bogolons near the Fermi level.
The enclosed regions with dotted lines show the low-energy windows relevant to instability of Bogoliubov Fermi surfaces.
}
\label{fig:band}
\end{center}
\end{figure}

Since the low-temperature behaviors are dominated by the degrees of freedom near the Fermi level, we can construct the effective low-energy model involving only the HBZ.
There are doubly degenerate components near the Fermi level protected by the particle-hole symmetry,
which are labeled as $1$ and $2$, and the corresponding effective low-energy Hamiltonian is written as
\begin{align}
\mathscr H_{\rm eff} &= \sum_{\bm k \in \Omega}
\Big( 
 \ep_{\bm k1} \al_{\bm k1}^\dg \al_{\bm k1}
+\ep_{\bm k2} \al_{\bm k2}^\dg \al_{\bm k2}
\Big)
,
\label{eq:ham_eff_1}
\end{align}
where $\ep_{\bm k1} = -\ep_{\bm k2} \equiv \ep_{\bm k}$.
The region $\Omega$ represents a wave-vector space near the Fermi surfaces.
Whereas the energy is dependent on the pseudospin index (1,2), it is more convenient to define another spinless fermion $\al_{\bm k \in \Omega} = \al_{\bm k1}$ and $\al_{\bm k \in \Omega'} = \al_{-\bm k,2}^\dg$ (see Fig.~\ref{fig:band} for schematic pictures of band structures and definition of $\Omega'$).
Then 
we obtain the simple spinless Hamiltonian
\begin{align}
\mathscr H_{\rm eff} &= \sum_{\bm k \in \Omega+\Omega'}
\ep_{\bm k} \al^\dg_{\bm k} \al_{\bm k}
,
\label{eq:ham_eff_2}
\end{align}
where the constant term is dropped.
We have used the relation $\ep_{-\bm k} = \ep_{\bm k}$.
Since one sees the unclosed Fermi-surfaces for $\bm k \in {\rm HBZ}$ in Eq.~\eqref{eq:ham_eff_1}, 
the Hamiltonian \eqref{eq:ham_eff_2} is more natural representation where only closed Fermi surfaces exist.
We have the inversion symmetry in Eq.~\eqref{eq:ham_eff_2}, which is seen as the particle-hole symmetry in terms of Eq.~\eqref{eq:ham_eff_1}.

\subsection{Order parameter and susceptibility}

\begin{figure}[t]
\begin{center}
\includegraphics[width=75mm]{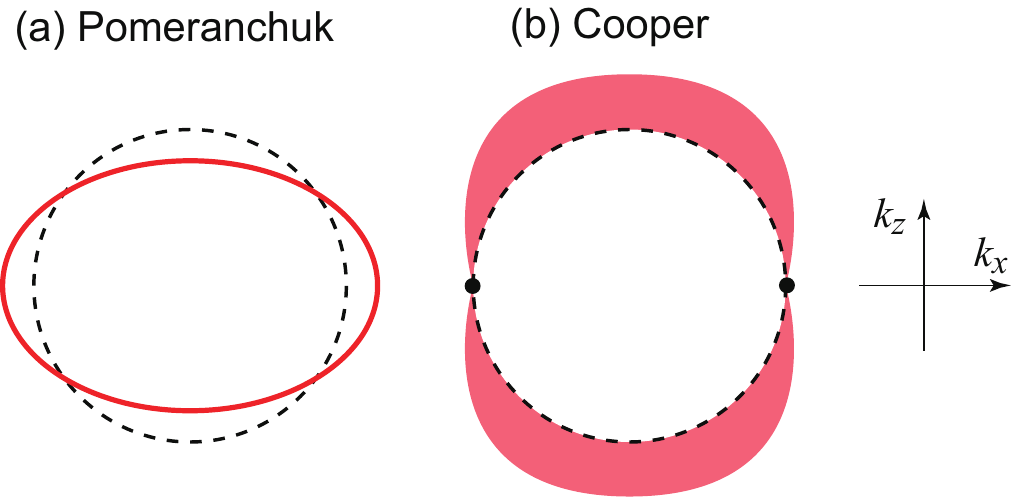}
\caption{
Schematic pictures for (a) diagonal and (b) offdiagonal orders of $\eta = z$ type [$g_z(\bm k) \propto k_z$].
The dotted line shows the bare Fermi surface without ordering.
The solid line in (a) shows the deformed Fermi surface due to the Pomeranchuk instability.
The shaded area in (b) shows the magnitude of gap induced from the Cooper instability.
}
\label{fig:insta}
\end{center}
\end{figure}

With the Hamiltonian \eqref{eq:ham_eff_2}, the analogy to spinless fermions can be used and now we are ready to define the possible order parameters of bogolons.
We have two kinds of order parameters, i.e., diagonal (Pomeranchuk instability) and offdiagonal (Cooper instability) ones.
The former is defined by
\begin{align}
\mathscr N_{\rm D,\eta} &= \sum_{\bm k \in \Omega+\Omega'} g_\eta(\bm k) \al_{\bm k}^\dg \al_{\bm k}
,
\label{eq:def_diag}
\end{align}
where the symbol `$\mathrm D$' represents a diagonal component.
$\eta$ represents polynomials such as e.g. $g_{\eta=xy}(\bm k) \propto k_x k_y$ (see Appendix B for more details), which determines the spatial structure.
The instability toward these orders can be studied based on the susceptibility defined by
\begin{align}
\chi_{\mathrm D,\eta} &= \int_0^{1/T} \diff \tau \, \la \mathscr N_{\rm D,\eta} (\tau) \mathscr N_{\rm D,\eta}^{\dg} \ra
.
\label{eq:chi_diag}
\end{align}
The Heisenberg picture with imaginary time is defined as $\mathscr O(\tau) = \epn^{\tau \mathscr H_{\rm eff}} \mathscr O \epn^{-\tau \mathscr H_{\rm eff}}$, and
the bracket $\la \cdots \ra$ means the quantum statistical average using $\mathscr H_{\rm eff}$.
The physical meaning of this quantity is how much of 
the order parameter is induced as $\la \mathscr N_{\rm D,\eta} \ra = \chi_{\rm D, \eta} h_{\rm D,\eta}$ under the test field $h_{\rm D,\eta}$ conjugate to $\mathscr N_{\rm D,\eta}$. 
The schematic picture of Pomeranchuk instability is shown in Fig.~\ref{fig:insta}(a).
As shown later, this is simply evaluated to give $\chi_{\rm D,\eta}\sim \rho$ with $\rho$ being a density of states at the Fermi level.
If the repulsive interaction $U>0$ is present, this susceptibility is enhanced by the factor $(1-U\chi_{\rm D,\eta})^{-1}$ according to the random phase approximation.
Hence the ordering occurs when the Stoner condition $\rho U\gtrsim 1$ is satisfied.

At this point, we introduce the matrix $\hat O^{\eta'}$ related to the multipoles where $\eta'$ again represents polynomial functions.
The full functional forms are given in Appendix A, which are classified into
one monopole ($\hat O^1 \sim \hat 1$ with the identity $\hat 1$), 
three dipoles (e.g. $\hat O^x = \hat J^x$),
five quadrupoles (e.g. $\hat O^{xy} \sim \hat J^x \hat J^y$), and
seven octupoles (e.g. $\hat O^{xyz} \sim \hat J^x \hat J^y \hat J^z$).
These are called electronic multipoles in this paper.
With the above 16 matrices, any $4\times 4$ matrices can be 
expanded.
For example, the spin-orbit interaction term with $\beta$ [Eq.~\eqref{eq:ham_orig}] is expressed by the quadrupoles.
We can also classify the pairing amplitudes by the matrix $\hat O^{\eta'}\hat E$, as
spin-singlet ($\eta'=1$),
spin-triplet (e.g. $\hat O^{x} \hat E$),
spin-quintet (e.g. $\hat O^{xy} \hat E$), and
spin-septet  (e.g. $\hat O^{xyz} \hat E$),
which are called electronic multiplet pairs.
Our choice of the pair potential in Eq.~\eqref{eq:def_Delta} is regarded as a mixture of spin-singlet and spin-quintet \cite{Agterberg17}.

It is interesting to see the bogolons' physical quantities in terms of the {\it original electrons}.
The operator for the bogolons' order parameter is expanded by the multipole operators of electrons as
\begin{align}
\mathscr N_{\rm D,\eta} &= 
\sum_{\bm k \in \Omega}\sum_{\eta'}\sum_{\xi'={\rm D},\bar {\rm D},{\rm O},\bar {\rm O}}
C_{\rm D, \eta}^{\xi'\eta'} (\bm k) 
\vec \psi_{\bm k}^\dg 
\check O^{\xi'\eta'}
\vec \psi_{\bm k}
+{\rm const.}
\\
&\equiv \sum_{\xi'\eta'} \mathscr M_{\rm D,\eta}^{\xi'\eta'}
+{\rm const.}
\label{eq:multipole1}
\end{align}
The $8\times 8$ matrix $\check O^{\xi'\eta'}$ composed of $\hat O^{\eta'}$ and $\hat E$ specifies the type of electron multipoles defined in Appendix A3.
Equation \eqref{eq:multipole1} shows that the operator $\mathscr N_{\rm D}$ for bogolons
includes diagonal component $\mathscr M^{\xi'={\rm D},\bar {\rm D}}$ where ${\rm D}$ and $\bar {\rm D}$ represent even-parity and odd-parity multipoles.
These two types ($\xi'={\rm D},\bar {\rm D}$) can be constructed due to the enlargement of the single-particle Hilbert space involving $\bm k$ and $-\bm k$ components.
Note that the internal $j=3/2$ degrees of freedom are ragarded as independent from the parity here ($\bm k$-parity).
We also have the offdiagonal components ${\mathscr M}^{\xi'={\rm O},\bar {\rm O}}$ of original electrons due to the gauge symmetry breaking by $\hat \Delta_{\bm k}$, where the symbol ${\rm O}$ shows electron pair ($c^\dg c^\dg$-type) and $\bar {\rm O}$ shows hole pair ($c c$-type).
The expansion coefficient $C$ in Eq.~\eqref{eq:multipole1} is to be evaluated in the next subsection.

The above discussion is also extended to the pairing state of bogolons \cite{Oh19}:
\begin{align}
\mathscr N_{\rm O,\eta} &= \sum_{\bm k \in \Omega+\Omega'} g_\eta(\bm k) \al_{\bm k}^\dg \al^\dg_{-\bm k}
,
\label{eq:def_pair}
\end{align}
where the symbol `${\rm O}$' in the left-hand side represents an offdiagonal component (Cooper pairing).
Since the present Bogoliubov Fermi surfaces do not have internal degrees of freedom, the form factor must satisfy $g_{\eta}(- \bm k) = - g_\eta(\bm k)$ (odd parity) due to the Pauli principle.
The multipole expansion is performed as
\begin{align}
\mathscr N_{\rm O,\eta} &= 
\sum_{\bm k \in \Omega}\sum_{\xi'\eta'}
C^{\xi'\eta'}_{\rm O,\eta} (\bm k) 
\vec \psi_{\bm k}^\dg 
\check O^{\xi'\eta'}
\vec \psi_{\bm k}
\\
&\equiv 
\sum_{\xi'\eta'} \mathscr M^{\xi'\eta'}_{\mathrm O, \eta}
.
\label{eq:multipole2}
\end{align}
The non-interacting pair susceptibility $\chi_{\rm O,\eta}$, which is defined in a manner similar to Eq.~\eqref{eq:chi_diag}, is evaluated to give $\chi_{\rm O,\eta} \sim \rho \ln (\omega_c/T)$ with $\omega_c$ being a cutoff energy.
When the attractive interaction $U<0$ acts on bogolons, the pairing susceptibility is enhanced by the factor $(1+U\chi_{\rm O,\eta})^{-1}$, and the logarithmic divergence gives rise to the pairing state below the finite transition temperature $T_c \sim \omega_c \exp (-\frac 1 {\rho |U|} )$.
The schematic picture of the Cooper instability is shown in Fig.~\ref{fig:insta}(b).

\begin{table*}
\caption{
List of meaning of the susceptibility tensor $\chi_{\xi\eta}^{\xi'\eta'}$, where (a) the index $(\xi,\eta)$ represents information for bogolons and (b) the other one $(\xi',\eta')$ for the original $j=3/2$ electrons.
}
\label{tab:1}
\vspace{2mm}
\begin{tabular}{|c|c|c|}
\multicolumn{3}{l}{(a) {\it Operators for bogolons} $\mathscr N_{\xi\eta}$} \\
\hline
\multirow{2}{*}{\ $\xi$\ } & ${\rm D}$ & Diagonal (Pomeranchuk) \\
\cline{2-3}
                        & ${\rm O}$ & Offdiagonal (Cooper) \\
\hline\hline
\multirow{7}{*}{\ $\eta$\ } & $1$ & $s$-wave (even-parity) \\
\cline{2-3}
                        & $x,y,z$ & $p$-wave (odd-parity) \\
\cline{2-3}
                        & $xy,yz,zx,3z^2-r^2,x^2-y^2$ & $d$-wave (even-parity) \\
\cline{2-3}
                        & $xyz, x(5z^2-r^2), y(5z^2-r^2)$ & \multirow{3}{*}{$f$-wave (odd-parity)} \\
                        & $z(5z^2-3r^2), z(x^2-y^2)$ & \\
                        & $x(x^2-3y^2), y(3x^2-y^2)$ & \\
\cline{2-3}
                        & $\vdots$ & $g$-wave, $\cdots$ \\
\hline
\end{tabular}
\hspace{2mm}
\begin{tabular}{|c|c|c|}
\multicolumn{3}{l}{(b) {\it Operators for $j=3/2$ electrons} $\mathscr M_{\xi\eta}^{\xi'\eta'}$} \\
\hline
\multirow{2}{*}{\ $\xi'$\ } & ${\rm D}$/$\bar{\rm D}$ & Even/Odd-parity multipole \\
\cline{2-3}
                        & ${\rm O}$/$\bar{\rm O}$ & Electron/Hole multiplet pair \\
\hline\hline
\multirow{7}{*}{\ $\eta'$\ } & $1$ & monopole or singlet \\
\cline{2-3}
                        & $x,y,z$ & dipole or triplet \\
\cline{2-3}
                        & $xy,yz,zx,3z^2-r^2,x^2-y^2$ & quadrupole or quintet \\
\cline{2-3}
                        & $xyz, x(5z^2-r^2), y(5z^2-r^2)$ & \multirow{3}{*}{octupole or septet} \\
                        & $z(5z^2-3r^2), z(x^2-y^2)$ & \\
                        & $x(x^2-3y^2), y(3x^2-y^2)$ & \\
\hline
\multicolumn{3}{l}{\vspace{2.85mm}} \\
\end{tabular}
\end{table*}

Summing up the above expressions, we can rewrite the linear response formula 
$\la \mathscr N_{\xi\eta} \ra = \chi_{\xi\eta} h_{\xi\eta}$ as
\begin{align}
\la \mathscr M_{\xi\eta}^{\xi'\eta'} \ra &= \chi_{\xi\eta}^{\xi'\eta'} h_{\xi\eta}
\label{eq:lin_resp_detail}
\\
\chi_{\xi\eta}^{\xi'\eta'} &= 
\int_0^{1/T} \diff \tau \, \la \mathscr M^{\xi'\eta'}_{\xi\eta} (\tau) \mathscr N_{\xi\eta}^{\dg} \ra
\end{align}
with the sum rule 
\begin{align}
\chi_{\xi\eta} = \sum_{\xi'\eta'}\chi_{\xi\eta}^{\xi'\eta'}
.
\label{eq:sum_orig}
\end{align}
The physical meaning of $\chi_{\xi\eta}^{\xi'\eta'}$ in Eq.~\eqref{eq:lin_resp_detail} is as follows: once the internal test field $h_{\xi\eta}$ corresponding to the diagonal($\xi={\rm D}$)/offdiagonal($\xi={\rm O}$) ordering of bogolons, the conjugate order parameter $\la \mathscr N_{\xi\eta} \ra$ of bogolons is induced simultaneously, which is composed of the multipoles ($\xi'={\rm D},\bar{\rm D}$) and multiplet pairs ($\xi'={\rm O},\bar {\rm O}$) of original electrons with a type $\eta'$ (see Tab.~\ref{tab:1}).
Simply speaking, the susceptibility gives the information on what types of electrons' multipole/Cooper pairs are induced by the bogolons' ordering.
While we have a number of susceptibilities, in our model we can utilize a continuous rotational symmetry in the $xy$-plane and some components are found to be identical with each other.

To avoid confusions, we note the rules of the notations in this paper: the symbols $(\xi,\eta)$ and the `prime' ones $(\xi',\eta')$ in the quantity $\chi_{\xi\eta}^{\xi'\eta'}$ respectively refer to the bogolons and original $j=3/2$ electrons.
For example, let us consider $\chi_{\xi=\mathrm D, \eta=x}^{\xi'=\mathrm D, \eta'=xy}$.
This quantity tells us how much of the quadrupole of $\hat O^{xy}$ ($\sim \hat J_x \hat J_y$) type  is induced from the ordering of $p$-wave type diagonal order (or sometimes called electron nematic state) of bogolons.
For $\chi_{\xi=\mathrm O, \eta=x}^{\xi'=\mathrm O, \eta'=xyz}$, on the other hand, it means how much of the spin-septet pairs (with odd parity) is induced from the $p$-wave pairing state of bogolons.
The other combinations are also interpreted in a similar manner.

\subsection{Evaluation of susceptibility tensor}
Here we show the method how to calculate the susceptibility introduced in the last subsection.
The order parameter of bogolons can be written in the form
\begin{align}
\mathscr N_{\xi\eta} &= \sum_{\bm k\in \Omega} g_\eta (\bm k) \vec \al_{\bm k}^\dg \check n_{\xi\eta} \vec \al_{\bm k}
.
\end{align}
Note that the wavevector summation is performed within $\Omega$ (see Fig.~\ref{fig:band}).
The matrix elements are defined by
\begin{align}
\check n_{\rm D\eta} &=
\begin{pmatrix}
1&&&&&&&\\
&-(-1)^\eta&&&&&&\\
&&0&&&&&\\
&&&0&&&&\\
&&&&0&&&\\
&&&&&0&&\\
&&&&&&0&\\
&&&&&&&0\\
\end{pmatrix}
\end{align}
and
\begin{align}
\check n_{\rm O\eta} &=
\begin{pmatrix}
0&1-(-1)^\eta&&&&&&\\
0&0&&&&&&\\
&&0&&&&&\\
&&&0&&&&\\
&&&&0&&&\\
&&&&&0&&\\
&&&&&&0&\\
&&&&&&&0\\
\end{pmatrix}
,
\end{align}
where the first and second rows/columns indicate the Bogoliubov particles that have Fermi surfaces, and the others are away from the Fermi level. 
The sign is defined by the parity relation $g_\eta (-\bm k) = (-1)^\eta g_\eta (\bm k)$.
The multipole expansions in Eqs.~\eqref{eq:multipole1} and \eqref{eq:multipole2} are explicitly performed as
\begin{align}
\mathscr M_{\xi\eta}^{\xi'\eta'}
 &= \sum_{\bm k} C_{\xi\eta}^{\xi'\eta'}(\bm k) 
\vec \psi_{\bm k}^\dg \check O^{\xi'\eta'} \vec \psi_{\bm k}
\\
C_{\xi\eta}^{\xi'\eta'}(\bm k)
&= 
g_\eta (\bm k) \, \trace \Big(
 \check O^{\xi'\eta'\dg} \check U_{\bm k} \check n_{\xi\eta} \check U^\dg_{\bm k}
\Big)
.
\end{align}
See Appendix A3 for the definition of the $8\times 8$ matrices.
We then obtain the susceptibility
\begin{align}
\chi_{\xi\eta}^{\xi'\eta'}
&= 
-T
\sum_{n\bm k}
g_\eta (\bm k)^2
\trace\big(
\check O^{\xi'\eta'\dg} \check U_{\bm k} \check n_{\xi\eta} \check U_{\bm k}^\dg
\big)
\nonumber \\
&\ \ \ 
\times \trace\Big[
\check U^\dg_{\bm k}
\check O^{\xi'\eta'}
\check U_{\bm k}
\check G_{\bm k}(\imu \omega_n)
\check n^{\dg}_{\xi\eta}
\check G_{\bm k}(\imu \omega_n)
\Big]
,
\end{align}
where we have defined the Green function
\begin{align}
\check G_{\bm k}(\imu\omega_n) &= \left(
\imu\omega_n \check 1 - \check \Lambda_{\bm k}
\right)^{-1}
\label{eq:green}
\end{align}
with the fermionic Matsubara frequency $\omega_n = (2n+1)\pi T$.
Let us make a brief comment on the structure of the frequency dependence:
if one stops at this expression and looks at the frequency structure,
one can study the contributions of even- and odd-frequency multiplet pairs \cite{Tanaka12}, although we focus on the
static properties
in this paper.
Then the odd-frequency (even-frequency) components appears together with even-parity (odd-parity) components due to the Pauli principle.
More generically, the expansion in the four-dimensional space-time is possible including both the diagonal and offdiagonal components \cite{Hoshino19}.

The Matsubara summation is performed and we obtain the simpler expression
\begin{align}
\chi_{\xi\eta}^{\xi'\eta'}
&= 
\sum_{\bm k} P_\xi (\ep_{\bm k})
\left|
C_{\xi\eta}^{\xi'\eta'}(\bm k)
\right|^2
\label{eq:simple}
,
\end{align}
where the energy dependent function is given by
\begin{align}
& P_{\xi={\rm D}}(\ep) =  - \frac{\partial f(\ep)}{\partial \ep}
, \\
& P_{\xi={\rm O}}(\ep) = 
-\frac{f(\ep) - f(-\ep)}{2\ep}
.
\end{align}
We have introduced the Fermi distribution function $f(\ep) = 1 / (\epn^{\ep/T}+1)$.

Now we perform the $\bm k$-integral.
Since our system has rotational symmetry around $z$-axis, the cylindrical coordinate $\bm k = (k, \phi, k_z)$ with $k_x = k\cos \phi$ and $k_y = k\sin \phi$ is convenient.
The unitary matrix that diagonalizes the original Hamiltonian is transformed by the rotation along $k_z$-axis as
\begin{align}
\check U_{\bm k} &= \check U(k,\phi,k_z)
= \check R (\phi) \check U(k, 0, k_z)
, \\
\check R(\phi) &=
\begin{pmatrix}
\exp \big[- \imu (\hat J_z - \tfrac 1 2 \hat 1) \phi \big] & \hat 0 \\
\hat 0 & \exp \big[\imu (\hat J_z - \tfrac 1 2 \hat 1) \phi \big]
\end{pmatrix}
.
\end{align}
The matrix $\check R$ represents a rotation around $k_z$-axis, 
which connects different $\bm k$ points in $xy$-plane.
Since the form factor $g_\eta$ is written as $g_\eta(\bm k) = g'_\eta(k,k_z) f_\eta(\phi)$ (see Appendix B for its concrete form),
the $\phi$-integral in evaluating susceptibility can be performed analytically, and the other integrals are considered within $k_x$-$k_z$ plane.
It is also convenient to change the coordinate system as $(k, k_z) \to (k_\perp, k_\parallel)$ which is the $\bm k$-coordinate normal and parallel to the Fermi surface.
The coordinate $k_\perp$ can be changed into energy integral as $\diff \ep = v \diff k_\perp$ where $v$ is the Fermi velocity. 
After some calculations, one reaches the expression
\begin{align}
\chi_{\xi\eta}^{\xi'\eta'}
&\simeq 
\frac{1}{(2\pi)^2}\int_{-\omega_c}^{\omega_c}
 \diff \ep P_\xi(\ep)
\int_{k_x,k_z>0} \diff k_\parallel 
\frac{k_x(k_\parallel) |g'_\eta(k_\parallel)|^2}
{|v(k_\parallel)|}
\nonumber \\
&\ \ \ \times \sum_{ijkl}
O_{ij}^{\xi'\eta'\dg}
\Big[
\check U (k_\parallel)
 \check n_{\xi\eta} 
\check U^\dg(k_\parallel)
\Big]_{ji}
\Gamma^\eta_{ijkl}
\nonumber \\
&\hspace{10mm} \times
O_{kl}^{\xi'\eta'}
\Big[
\check U (k_\parallel)
\check n_{\xi\eta}^\dg
\check U^\dg(k_\parallel) 
\Big]_{lk}
,
\label{eq:final_express}
\end{align}
where we have introduced the cutoff energy $\omega_c$ and have simplified the expression by assuming that the dominant contribution comes from the Fermi surface, i.e. $\ep = 0$.
The $\phi$-integral part has been separately given as
\begin{align}
&\Gamma^\eta_{ijkl}
= \int \frac{\diff\phi}{2\pi}
|f_\eta(\phi)|^2
\epn^{\imu (n_i-n_j+n_k-n_l)\phi}
,
\end{align}
where the integer $n_i$ is defined by $R_{ii}(\phi) = \epn^{-\imu n_i \phi}$.
The function $\Gamma_{ijkl}^\eta$ can be evaluated by using the information on Appendix B.
Thus we only have to evaluate the $k_\parallel$ line integral along the Fermi surface, which is performed numerically with the discretized mesh.
Since the wavevector $\bm k$ belongs to $\Omega$,
the $k_\parallel$-integral is performed within the region of $k_x,k_z>0$.

\subsection{Interpretation in terms of Landau theory}

Before showing the numerical results,
let us discuss the Landau free energy relevant to Pomeranchuk/Cooper instabilities of bogolons 
to understand the induced electron multipoles and multiplet pairs. 
We consider the one of the bogolon orderings whose order parameter is denoted as $n$ \big($=\la \mathscr N_{\xi\eta} \ra$\big) where we have omitted the index $(\xi,\eta)$ for simplicity.
The corresponding physical quantities for original electrons are written as $m_i$ \big($\sim \la \mathscr M_{\xi\eta}^{\xi'\eta'} \ra$\big) where we have used the short-hand notation $i=(\xi'\eta')$.
Since some of the electron multipoles are finite from the biginning in our model, $m_i$ is defined as a deviation from its equilibrium point.
The Landau free energy is explicitly written down as
\begin{align}
F &= a n^2 + b n^4 - h_{\rm ext} n + \sum_i g_i n m_i + \sum_i a_i' m_i^2
.
\end{align}
$h_{\rm ext}$ is an imaginary test field for bogolons, and $g_i$ is the coupling between bogolon's and electrons' order parameters.
We assume $a_i'>0$ and $b>0$ which guarantees the thermodynamic stability.

The first term in the right-hand side represents a distance to the critical point, and is rewritten as
\begin{align}
an^2 &= a_0 n^2 - h_{\rm MF}(n) n
,
\label{eq:gl_mf}
\end{align}
where $a_0^{-1}$ ($>0$) is a free susceptibility without interactions, and $h_{\rm MF} = I n$ is the internal mean-field induced from an effective interaction $I$.
In order to have an intuition of Eq.~\eqref{eq:gl_mf}, let us consider the two examples.
First, if the non-interacting susceptibility shows Curie law as $a_0(T)^{-1} \sim \frac 1 T$, we have in the presence of interaction the Curie-Weiss law $a(T)^{-1} \sim \frac{1}{T-T_c}$ with the Curie temperature $T_c \sim I$ determined by the condition $a=0$.
Second, if we consider a pairing state of electrons, we have $a_0(T)^{-1} \sim \rho \ln \frac{\omega_c}{T}$ from the Fermi-surface instability.
Then the full inverse susceptibility is given by $a(T) \sim \rho I^2 \ln \frac{T}{T_c}$ with the superconducting transition temperature $T_c \sim \omega_c \exp (- \frac{1}{\rho I})$, which corresponds to a standard result of the BCS theory.
We note that, as seen below, the transition temperature is modified once the order parameter $n$ couples to the other ones ($m_i$) linearly.

Above the transition temperature of the bogolon's ordering,
the order parameters $n$, $m_i$ are just induced from the external field.
The resultant order parameters are given by
\begin{align}
n &= \chi h_{\rm ext}
\\
m_i &= f_i n = f_i \chi h_{\rm ext}
,
\end{align}
where the susceptibility $\chi$ and the factor $f_i$ are
\begin{align}
\chi &= \left( 
2a - \sum_i 2a_i' f_i^2
 \right)^{-1}
,
\\
f_i &= - \frac{g_i}{2a_i'}
,
\end{align}
According to Eq.~\eqref{eq:sum_orig}, there is the sum rule $n=\sum_i m_i$, or
\begin{align}
\sum_i f_i = 1
.
\label{eq:sumrule}
\end{align}
Thus the magnitude of electron order parameter induced from external field for bogolons is controlled by the factor $f_i$.
What we have calculated in Eq.~\eqref{eq:final_express} of the main text is the quantity $f_i$.
If one 
traces out $m_i$ from the equations of state, only the order parameter $n$ enters to the Landau theory and the coefficient $a$ is replaced by $\chi^{-1}/2$.
The $n$-only model corresponds to the effective low-energy model in Eq.~\eqref{eq:ham_eff_2}, which does not include the information of original electrons.

Below the transition temperature determined by $\chi^{-1}=0$, on the other hand, the bogolon order parameter and corresponding electron order parameters are induced from the self-consistent internal field as
\begin{align}
n &= \frac{1}{I} h_{\rm MF}
,
\\
m_i &= f_i n = \frac{f_i}{I} h_{\rm MF}
,
\end{align}
where $1/I$ plays a role of susceptibility for the internal mean-field.
Note that the factor $f_i$ is the same as the one above the transition temperature.
Hence, if we would like to know the electrons' multipole components in the bogolons' ordered state, we only have to calculate the factor $f_i$ in the bogolons' disordered phase.
In the next section, we show the numerical results for $f_i$ calculated in the presence of Bogoliubov Fermi surfaces.

\section{Numerical results}

\subsection{Bogoliubov Fermi surfaces}

\begin{figure}[t]
\begin{center}
\includegraphics[width=75mm]{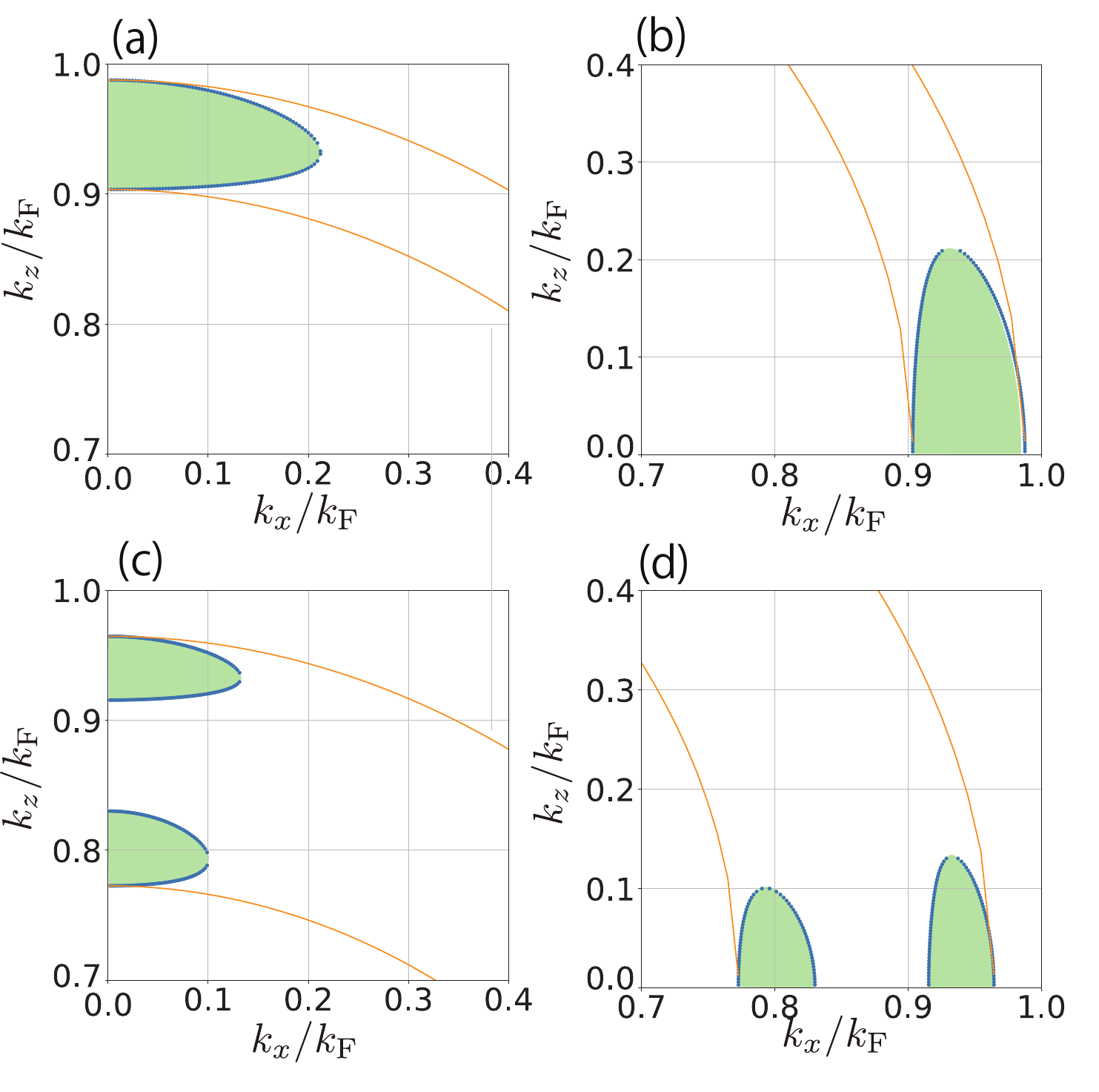}
\caption{
Shapes of Bogoliubov Fermi surfaces.
The parameters are chosen as $\beta/\ep_{\rm F} = 0.1$, $\Delta_0/\ep_{\rm F} = 0.1$, $\Delta_1/\ep_{\rm F} = 0.5$ for (a,b) and $\beta/\ep_{\rm F} = 0.3$, $\Delta_0/\ep_{\rm F} = 0.1$, $\Delta_1/\ep_{\rm F} = 0.5$ for (c,d).
The electron Fermi surfaces without pair potentials are drawn by yellow solid lines.
The blue dots  show the Bogoliubov Fermi surface with the enclosed Fermi volume shaded by green color.
}
\label{fig:fermi}
\end{center}
\end{figure}

Before showing the susceptibilities,
let us first discuss the Bogoliubov Fermi surfaces in our setup.
Figure~\ref{fig:fermi}(a,b) show the shapes of Fermi surfaces in the presence of the pair potential, where the parameters are chosen as $\beta = 0.1$, $\Delta_0 = 0.1$, $\Delta_1 = 0.5$ with the energy unit $\ep_{\rm F} = \al k_{\rm F}^2=1$.
In this case we have two Fermi surfaces in the first quadrant of $k_x$-$k_z$ plane, which are separately shown in (a) and (b). 
Depending on the choice of parameters, the number of the Fermi surfaces changes, and we can have four Fermi surfaces in total as shown in (c) and (d) for the parameters $\beta = 0.3$, $\Delta_0 = 0.1$, $\Delta_1 = 0.5$.
In fact, the shapes of the Fermi surfaces in (a) and (b) [also (c) and (d)] are same if they are inverted at $k_z=k_x$ line,
since the eigenenergies of $\check H(k_x,0,k_z)$ is identical to those of $\check H(k_z,0,k_x)$.

For comparison, the normal-state Fermi surfaces without pair potentials are also drawn with yellow lines.
We can see from  Figs.~\ref{fig:fermi}(a) and (b) that the Fermi surfaces near the $k_z$- and $k_x$-axes are not much modified, and away from the axes the deviation becomes larger.
The wave function on the Fermi surface also remains unchanged near the axes, i.e., no mixture between electrons and holes,
which 
is related to the disappearance of the pair potential proportional to $\Delta_1$ in $\Delta_{\bm k}$.

\begin{figure*}[t]
\begin{center}
\includegraphics[width=150mm]{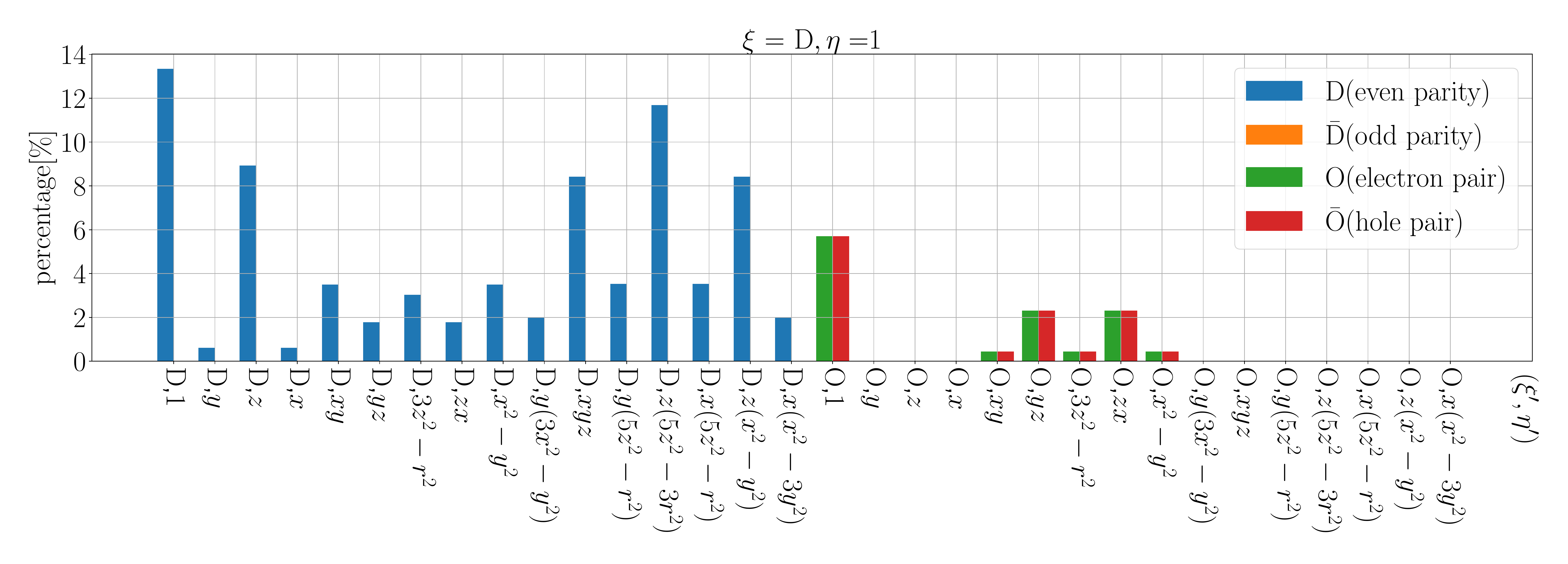}
\caption{
Electron multipole components $\chi_{{\rm D},1}^{\xi'\eta'}$ induced from the increased chemical potential for bogolons which form the Fermi surfaces.
The sum of the heights of bars is normalized to 100\%.
The parameters are chosen as $\beta/\ep_{\rm F} = 0.1$, $\Delta_0/\ep_{\rm F} = 0.1$, $\Delta_1/\ep_{\rm F} = 0.5$.
}
\label{fig:charge}
\end{center}
\end{figure*}

The green-shaded area enclosed by the Fermi surface is related to the three-dimensional Fermi volume if one rotates the plane around $z$-axis.
Then the Fermi volume near $k_z$-axis is pancake-like and the one near $k_x$-axis is donut-like \cite{Agterberg17}.
The Fermi surfaces around $k_x$-axis in Figs.~\ref{fig:fermi}(b,d) are away from $k_z$ axis, and the volume is much larger than the ones around $k_z$-axis.
The Fermi volume near $k_z$-axis is roughly ten times smaller than the one near $k_x$-axis.
Although the larger Fermi volume induces the larger instability toward ordering, it depends on the form factor $g_\eta(\bm k)$ which Fermi surface dominantly contributes.

We comment on another system that exhibits Bogoliubov Fermi surfaces.
The Bogoliubov Fermi surfaces inside the superconducting states are also proposed in the context of the Kondo lattices which is one of the basic models for heavy-electron materials \cite{Coleman93,Coleman94,Hoshino14PRL,Hoshino14PRB,Hoshino16}.
Here the origin of the peculiar superconductivity is the nontrivial effecitve hybridization between conduction and localized electrons, and the time-reversal symmetry breaking is not necessary for the mechanism.
These systems should also show the further ordering instabilities similar to those discussed in our paper.

\subsection{Susceptibility tensors}

According to the results in the last section, the non-interacting susceptibility can be written as
\begin{align}
\chi_{\xi\eta}^{\xi'\eta'} (T) &=
\tilde \chi_{\xi\eta}^{\xi'\eta'}
Q_\xi (T)
,
\label{eq:suscep_coef}
\end{align}
where $Q_{\rm D}(T) = 1$ and $Q_{\rm O}(T) = \ln \left( 2\epn^\gamma \omega_c / \pi T \right)$ with the Euler's constant $\gm\simeq 0.577$.
Since the temperature dependent part is not affected by the choice for $\eta,\eta'$,
we only have to numerically evaluate the temperature-independent coefficient $\tilde \chi_{\xi\eta}^{\xi'\eta'}$ ($>0$). 
The overall tendency of these physical quantities is not sensitive to the choice of the parameters.
Hence, below we concentrate on the results for the parameters $\beta=0.1$, $\Delta_0=0.1$, $\Delta_1=0.5$  whose Fermi surfaces are shown in Fig.~\ref{fig:fermi}(a,b).
The information on Fermi surfaces are extracted by dividing $k_x$-$k_z$ plane around the Fermi surface with $100\times 400$ grids and by using the linear interpolation.
We have checked that the quantitatively same results are obtained for the
finer mesh.

Figure~\ref{fig:charge} shows the normalized susceptibility coefficient $\tilde \chi_{\xi\eta}^{\xi'\eta'}$ defined in Eq.~\eqref{eq:suscep_coef}.
Here $\xi={\rm D}$ and $\eta=1$ ($s$-wave) is chosen for the bogolon's component, and the values are normalized by the constant $C_{\xi\eta} = \sum_{\xi'\eta'} \tilde \chi_{\xi\eta}^{\xi'\eta'}$ to be 100\percent\ in total.
This ($\xi={\rm D},\eta=1$) component means the deviation induced from the symmetric chemical potential field for bogolons, which does not break the symmetry of the original Hamiltonian in Eq.~\eqref{eq:ham_eff_2}.
Hence the finite values in Fig.~\ref{fig:charge} indicate the components that are originally finite without Pomeranchuk/Cooper instabilities.
Note that some components (e.g., $\eta'=x$ and $\eta'=y$) have the same value due to the symmetry.

One may have an impression that the finite $\xi'=x$ component ($\sim \hat J_x$) breaks the rotational symmetry around $z$-axis and is not consistent with the disordered situation.
This discrepancy at first sight is rationalized by noticing the fact that the multipole component in Fig.~\ref{fig:charge} concerns the internal degrees of freedom of $j=3/2$ electron.
The quantity is evaluated by the $\bm k$ integral in Eq.~\eqref{eq:simple} and the $\bm k$-dependence does not explicitly appear.
Hence, if one considers the multipole expansions in $\bm k$-space, the combination between $(k_x,k_y,k_z)$ and $\hat J_x$ can lead to a scalar component which does not break the original symmetry.
One can also show that the $\xi'=x$ component, which is absent in the original Hamiltonian, is induced in the Green function in Eq.~\eqref{eq:green} which is more directly connected to the physical quantities.
On the other hand, the odd-parity component and the spin-triplet/septet pair amplitudes cannot appear with a scalar form.
Then one needs the spontaneous symmetry breaking either by the Pomeranchuk or Cooper instability.

\begin{figure*}
\begin{center}
\includegraphics[width=150mm]{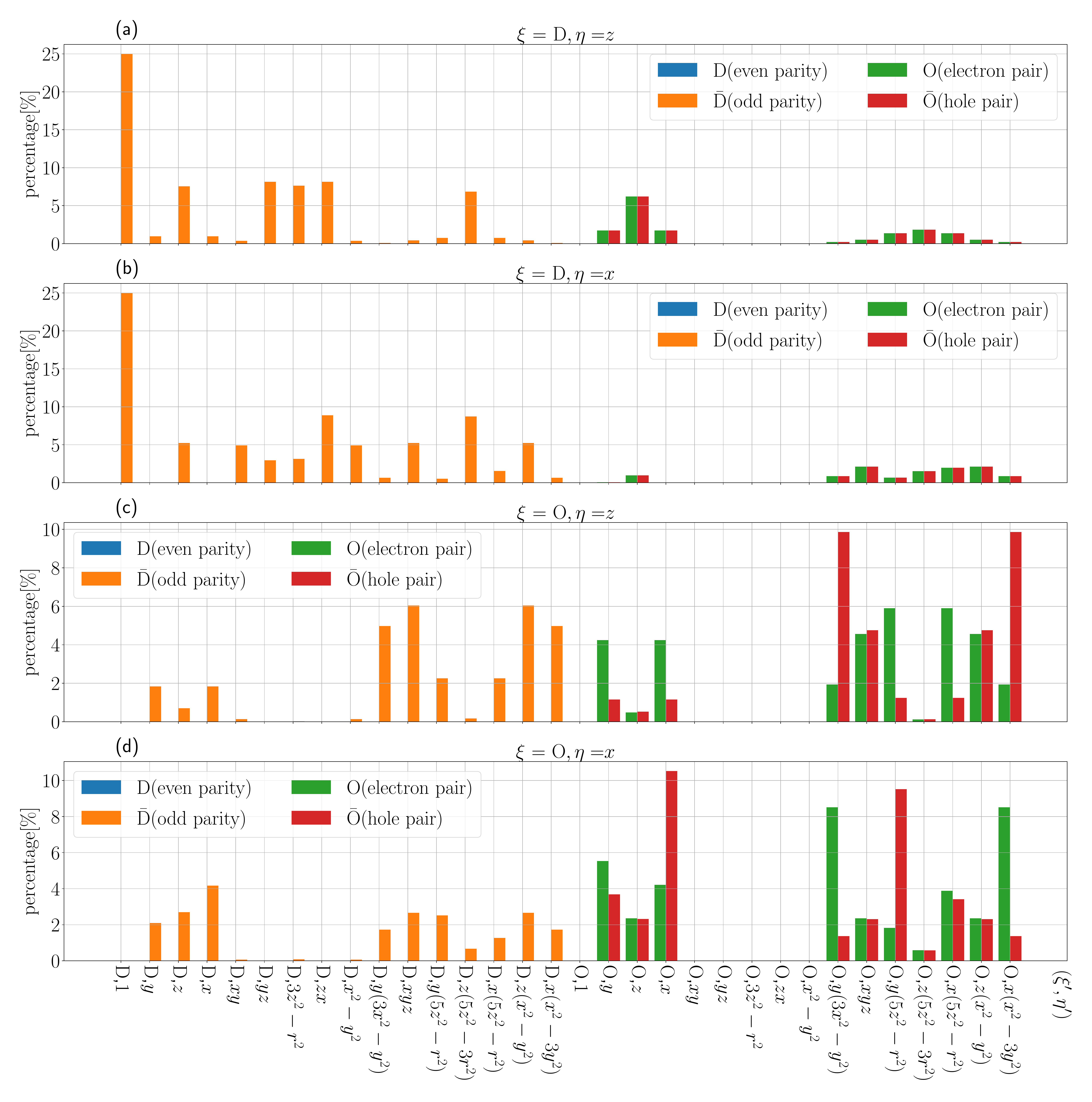}
\caption{
Electron multipole components induced from $p$-wave (a,b) Pomeranchuk ($\xi={\rm D}$) and (c,d) Cooper ($\xi={\rm O}$) instabilities of bogolons with the form factors (a,c) $g_{\eta=z}(\bm k) \propto k_z$ and (b,d) $g_{\eta=x}(\bm k) \propto k_x$.
The parameters are chosen as $\beta/\ep_{\rm F} = 0.1$, $\Delta_0/\ep_{\rm F} = 0.1$, $\Delta_1/\ep_{\rm F} = 0.5$.
}
\label{fig:pwave}
\end{center}
\end{figure*}

Next we discuss the results for symmetry-broken states of bogolons.
Since the results for $p$-wave ordering are similar to those for $f$-wave ordering, here we focus on the case with $\eta=z,x$.
Although we can also consider the $d$-wave components for the Pomeranchuk instability as in Tab.~\ref{tab:1}(a), the results are similar to the $s$-wave case in Fig.~\ref{fig:charge}.
This is because the $\bm k$-dependences do not explicitly enter in our multipole expansion as discussed above.
For the type of $\eta=xy$, for example, we have the $k_x k_y$-type order parameter within the charge sector ($\xi'={\rm D}$, $\eta'=1$), which is zero in the disordered phase but finite below $T_c$.
However, this type of order parameter cannot be explicitly seen in the figure where one only sees the $\bm k$-summed multipole expansion coefficients with respect to the internal degrees of freedom of $j=3/2$ electrons.
In order to look at the symmetry breakings of $d$-wave type, the spectral decomposition in terms of the angle $\phi$-functions is necessary.
Moreover, for a material-specific case, the Fermi surfaces are in general anisotropic, and the classification based on irreducible representations are necessary \cite{Brydon18}.
This decomposition is in principle possible, but
here we focus on the typical cases of $p$-wave types to demonstrate the concept of the bogolons' orderings and their interpretation in terms of $j=3/2$ internal degrees of freedom of original electrons.

Figure \ref{fig:pwave} shows the results for the bogolons' ordering of $p$-wave types.
As shown in (a) and (b), which are plots for Pomeranchuk instabilities ($\xi={\rm D}$), all the components, i.e., odd-parity multipoles and triplet/septet pair amplitudes, are regarded as order parameters since these are zero in Fig.~\ref{fig:charge} without symmetry breaking.
Since the tendencies for $\eta=z$ and $\eta=x$ are similar,
let us take a close look at $\eta=z$ case in Fig.~\ref{fig:pwave}(a).
The magnitudes for diagonal and off-diagonal components are nearly $70$\percent\  and $30$\percent,
respectively.
Namely, the major part of the contributions are from diagonal ones.
Among them, the dominant one is $\eta'=1$ (monopole, or electron charge) component which occupies 25\percent.
Hence, whereas we consider the Pomeranchuk instability of bogolons, the dominant contribution is the same as that of electrons with monopole where internal degrees of freedom are not reflected.
Note that the other components also give contributions to the susceptibility although they are 
smaller than the monopole contribution.

Actually, the dominant component from $(\xi',\eta')=(\bar {\rm D},1)$ can be evaluated analytically, since the multipole matrix $\check O^{\xi'=\bar{ \rm D}, \eta'=1}$ is proportional to the identity matrix.
Then the expression for the susceptibility is substantially simplified and we obtain
\begin{align}
\chi_{\xi\eta}^{\bar{\rm D},1} = 
\frac{\trace \check n_{\xi\eta} \, \trace \check n^\dg_{\xi\eta}}{\trace [\check n_{\xi\eta}\check n^\dg_{\xi\eta}] \, \trace \check 1} 
\, \chi_{\xi\eta}
.
\end{align}
Hence the value of $25$\percent\ in Figs.~\ref{fig:pwave}(a) and (b) is the exact figure.
This expression also explains the disappearance of the electron monopole contribution for the offdiagonal bogolons' ordering discussed in the next, since the matrix $\check n_{{\rm O},\eta}$ is traceless.

In Figs.~\ref{fig:pwave}(c) and (d) we show the susceptibility tensor for the order parameters of electrons induced from Cooper instability of bogolons ($\eta=z,x$).
Since we expand the bogolon pair amplitude $\al_{\bm k}^\dg \al^\dg_{-\bm k}$, which is a non-Hermitian operator, the induced electron/hole pairs ($\xi'={\rm O},\bar{\rm O}$) have different magnitudes in contrast to Figs.~(a,b).
In the case of $\eta=z$ in (c), the magnitudes for diagonal and off-diagonal components are nearly $30$\percent\  and $70$\percent,
respectively, and main contributions come from offdiagonal ones.
The dominant contributions are the septet pairing with the types $\eta'=y(3x^2-y^2), x(x^2-3y^2)$ (each value is $\simeq 10$\percent).
However, these components are not as remarkable as the monopole in Figs.~\ref{fig:pwave}(a,b), and many pair amplitudes are induced simultaneously with comparable magnitudes.
The basic trends for the $\eta=x$ case in (d) are similar to the $\eta=z$ case, but the maximum contribution is the spin-triplet pair of $\eta'=x$ type.
Hence we conclude that the bogolons' Cooper pairs are dominantly contributed by the electron/hole pairs, while the main contribution of pair amplitudes are dependent on the types of bogolons' order parameters.
We note that the diagonal order parameters (electron multipoles) also give non-negligible contributions to bogolon Cooper pairs.
The similar features are seen also in $f$-wave pairs of bogolons.

In this way the bogolons' orderings partially inherit the properties of original electrons.
However, the bogolons' order parameter cannot be simply identified as one dominant component of electrons' order parameter.
Hence if one encounter the second phase transition inside the superconducting state, one cannot simply tell diagonal or offdiagonal order of original electrons, and we should consider the possibility of bogolon orders, where the electron diagonal/offdiagonal order components are substantially mixed with each other.
In connection to this point, the possible relevance to real materials is 
discussed in the next section.

\section{Summary and Discussions}

In this paper we have theoretically studied the possibility of the Pomeranchuk and Cooper instabilities of Bogoliubov Fermi surfaces below the superconducting transition temperature.
Using the $j=3/2$ electron model with symmetric spin-orbit interaction plus time-reversal symmetry broken even-parity pair potentials, we study the physics arising from remaining bogolons' degrees of freedom at low temperatures.
Based on the linear response theory, the bogolons' order parameters are systematically classified by using the multipole expansion both for diagonal and offdiagonal physical quantities of original electrons, i.e., multipoles and multiplet pairs.
These are also interpreted in the context of Landau free energy.

We numerically calculate the multipole expansion coefficients for the bogolons' order parameters.
For Pomeranchuk instability of bogolons, the main contribution comes from the monopole of electrons, and hence the Pomeranchuk instability of electrons' charge mainly occurs.
For Cooper instability of bogolons, the main contribution is the pair amplitudes of original electrons.
Thus the characters of electrons are partially inherited by bogolons.
We emphasize that the other minor components also give a non-negligible contribution to bogolon's order parameters.
Hence, below the superconducting transition temperature, there is no clear-cut answer in determining multipoles (diagonal) or multiplets (offdiagonal) as an order parameter for the second phase transition.

The bogolon orderings discussed in this paper can be a possible scenario for the materials which possesses a phase transition inside the superconducting phase.
In U$_{1-x}$Th$_x$Be$_{13}$ \cite{Smith85,Ott85,Heffner90, Shimizu17}, two-successive superconducting transitions have been observed.
The first transition at higher temperature can be identified as the superconducting order parameter, but the origin and the order parameter for the second transition have been argued.
The candidates are magnetic ordering and second superconducting order parameter with time-reversal symmetry breaking \cite{Heffner90,Shimizu17}.
In light of the present paper, we propose that the second transition cannot be simply classified into either diagonal (e.g. magnetic order) or offdiagonal (i.e. pairing state) orders in a strict sense, but rather they are mixed with each other in the presence of spin-orbital coupling.

On the other hand, in U$_{1-x}$Th$_x$Be$_{13}$ with applying pressure, whereas the second transition is not observed, the finite specific heat coefficient is found at low temperatures much lower than the transition point \cite{Zieve04}.
The remaining specific heat coefficient is also observed in the recently found superconductor UTe$_2$ \cite{Ran19,Aoki19}.
This indicates a remaining density of states even in the superconducting state.
Hence one expects some second orderings utilizing the degrees of freedom from Fermi surfaces at further low temperatures or by tuning the system.
Indeed, various superconducting phases recently identified in UTe$_2$ under the pressure and magnetic field \cite{Aoki20}.
Whereas ordinary superconductors do not have degrees of freedom below the transition temperature, the existence of the second phase transition indicates the presence of the remaining degrees of freedom which can be the existing Bogoliubov Fermi surfaces.

For another class of superconducting materials, recently the rotational symmetry breaking has been found in Bi$_2$Se$_3$- and iron-based materials \cite{Matano16, Yonezawa17, Pan16, Asaba17, Kushnirenko18} only in the superconducting state, which is called nematic superconductivity \cite{Yonezawa19}.
It has been proposed in Ref.~\cite{Fu14} that the superconducting state itself can produce the 
rotational symmetry breaking.
On the other hand, if the second transition is identified as it is separated from the first superconducting transition, there could be a possibility of ordering instability of remaining Fermi-surface degrees of freedom. 
Thus the insights of this paper will be useful for identifying the mechanisms and for analyzing properties of the superconductors which have second phase transitions.

\section*{Acknowledgement}
This work was supported by KAKENHI Grants 
No.~JP18K13490,
No.~JP18H01176,
No.~JP18H04305 and
No.~JP19H01842.

\appendix

\section{Multipole and multiplet-pair operators}

\subsection{Multipoles}

We consider the electronic multipole operator in Eq.~\eqref{eq:multipole1}, which is composed of 
\begin{align}
M^{\eta'}(\bm k) = \vec c_{\bm k}^\dg \hat O^{\eta'} \vec c_{\bm k}
.
\label{eq:multipole_def}
\end{align}
The matrices are defined by
\begin{align}
\hat O^{1} &= \sqrt{\tfrac 5 4} \, \hat 1
\end{align}
for monopole (charge),
\begin{align}
\hat O^{y} &= \hat J_y
= \frac {\imu} 2 
\begin{pmatrix}
0 & -\sqrt 3 &  & \\
\sqrt 3 & 0 & -2 & \\
& 2 & 0 & - \sqrt 3 \\
& & \sqrt 3 & 0
\end{pmatrix}
\label{eq:defJy}
\\
\hat O^{z} &= \hat J_z
= \frac 1 2 
\begin{pmatrix}
3 & & & \\
 & 1 & & \\
&  & -1 & \\
& &  & -3
\end{pmatrix}
\label{eq:defJz}
\\
\hat O^{x} &= \hat J_x
= \frac 1 2 
\begin{pmatrix}
0 & \sqrt 3 &  & \\
\sqrt 3 & 0 & 2 & \\
& 2 & 0 & \sqrt 3 \\
& & \sqrt 3 & 0
\end{pmatrix}
\label{eq:defJx}
\end{align}
for dipole,
\begin{align}
\hat O^{ xy} &= \sqrt{\frac 5 3} \lceil \hat J_x \hat J_y\rfloor 
\\
\hat O^{ yz} &= \sqrt{\frac 5 3} \lceil \hat J_y \hat J_z\rfloor 
\\
\hat O^{ 3z^2-r^2} &= \sqrt{\frac 5 {36}} (3\hat J_z^2-\hat{\bm J}^2)
\\
\hat O^{ zx} &= \sqrt{\frac 5 3} \lceil \hat J_z \hat J_x\rfloor 
\\
\hat O^{ x^2-y^2} &= \sqrt{\frac 5 {12}} (\hat J_x^2 - \hat J_y^2)
\end{align}
for quadrupole, and
\begin{align}
\hat O^{ y(3x^2-y^2)} &= \sqrt{\frac 5 {18}} \lceil \hat J_y (3\hat J_x^2 - \hat J_y^2)\rfloor 
\\
\hat O^{ xyz} &= \sqrt{\frac {20} 3} \lceil \hat J_x\hat J_y \hat J_z\rfloor 
\\
\hat O^{ y(5z^2-r^2)} &= \sqrt{\frac 5 {30}} \lceil \hat J_y (5\hat J_z^2 - \hat{\bm J}^2)\rfloor 
\\
\hat O^{ z(5z^2-3r^2)} &= \sqrt{\frac 5 {45}} \hat J_z (5\hat J_z^2 - 3\hat{\bm J}^2) 
\\
\hat O^{ x(5z^2-r^2)} &= \sqrt{\frac 5 {30}} \lceil \hat J_x (5\hat J_z^2 - \hat{\bm J}^2)\rfloor 
\\
\hat O^{ z(x^2-y^2)} &= \sqrt{\frac 5 3} \lceil \hat J_z (\hat J_x^2 - \hat J_y^2)\rfloor 
\\
\hat O^{ x(x^2-3y^2)} &= \sqrt{\frac 5 {18}} \lceil \hat J_x (\hat J_x^2 - 3\hat J_y^2)\rfloor 
\end{align}
for octupole, 
where the square bracket $\lceil \cdots \rfloor$ makes the operators symmetric and Hermitian as $\lceil ABC \rfloor = (ABC+ACB+BCA+BAC+CAB+CBA)/3!$, for example.
The above matrices are normalized as $\trace [(\hat O^{\eta'})^2] = 5$.

\subsection{Multiplet pairs}
We also define the antisymmetric tensor
\begin{align}
\hat E &= 
\begin{pmatrix}
&&&1\\
&&-1&\\
&1&&\\
-1&&&
\end{pmatrix}
\label{eq:anti_sym_tensor}
\end{align}
with which the pairing amplitude for electrons is defined by 
\begin{align}
P^{\eta'}(\bm k) = \vec c^\dg_{\bm k} \hat O^{\eta'}\hat E \vec c^{\dg{\rm T}}_{-\bm k}
.
\label{eq:multipole_def2}
\end{align}
Here the meaning of $\hat O^{\eta'}\hat E$ can be intuitively understood by comparing it with the two-body wave function composed of $j=3/2$ spins.
Let us consider the two-body wave function $|JM\ra$ ($M\in [-J,J]$) composed of two $j=3/2$ spins with the single-body wave function $|m\ra$ ($m\in [-j,j]$), which mimics the Cooper pair made of two electrons.
The wave function is classified by the total spin $J=0,1,2,3$, each of which corresponds to the spin-singlet, spin-triplet, spin-quintet and spin-septet states.
More specifically, we define the two-body wave function
\begin{align}
|\eta'\ra &= \sum_{mm'} (\hat O^{\eta'}\hat E)_{mm'}|m\ra_1 | m'\ra_2
\end{align}
and the full correspondence between the two-body states $|\eta'\ra$ and $|JM\ra$ is given as
\begin{align}
|\eta'=1\ra &\propto |0,0\ra
\end{align}
for spin-singlet state,
\begin{align}
|\eta'=y\ra &\propto |1,1\ra + |1,-1\ra
\\
|\eta'=z\ra &\propto |1,0\ra
\\
|\eta'=x\ra &\propto |1,1\ra - |1,-1\ra
\end{align}
for spin-triplet states,
\begin{align}
|\eta'=xy\ra &\propto |2,2\ra - |2,-2\ra
\\
|\eta'=yz\ra &\propto |2,1\ra + |2,-1\ra
\\
|\eta'=3z^2-r^2\ra &\propto |2,0\ra
\\
|\eta'=zx\ra &\propto |2,1\ra - |2,-1\ra
\\
|\eta'=x^2-y^2\ra &\propto |2,2\ra + |2,-2\ra
\end{align}
for spin-quintet states, and
\begin{align}
|\eta'=y(3x^2-y^2)\ra &\propto |3,3\ra + |3,-3\ra
\\
|\eta'=xyz\ra &\propto |3,2\ra - |3,-2\ra
\\
|\eta'=y(5z^2-r^2)\ra &\propto |3,1\ra + |3,-1\ra
\\
|\eta'=z(5z^2-3r^2)\ra &\propto |3,0\ra
\\
|\eta'=x(5z^2-r^2)\ra &\propto |3,1\ra - |3,-1\ra
\\
|\eta'=z(x^2-y^2)\ra &\propto |3,2\ra + |3,-2\ra
\\
|\eta'=x(x^2-3y^2)\ra &\propto |3,3\ra - |3,-3\ra
\end{align}
for spin-septet states.
These relations are also checked by the construction using Clebsch-Gordan coefficients.
Obviously, we can see the analogy to the $s,p,d,f$-electron's wave functions of the hydrogen atom written by polynomials of spatial coordinates.
For the electron pair amplitudes, we 
only have to 
replace $|m\ra_1$ by $c^\dg_{\bm km}$ and $|m'\ra_2$ by $c_{-\bm k,m'}^{\dg\rm T}$.

\subsection{Orthonormality}
With the above matrices, we define the $8\times 8$ matrices by
\begin{align}
\check O^{\mathrm D\eta'} &= 
\sqrt{\frac 1 {10}} 
\begin{pmatrix}
\hat O^{\eta'} & \hat 0 \\
\hat 0 & - \hat O^{\eta' {\rm T}}
\end{pmatrix}
\label{eq:1_of_4}
, \\
\check O^{\bar {\mathrm D}\eta'} &= 
\sqrt{\frac 1 {10}} 
\begin{pmatrix}
\hat O^{\eta'} & \hat 0 \\
\hat 0 & \hat O^{\eta' {\rm T}}
\end{pmatrix}
\label{eq:2_of_4}
, \\
\check O^{\mathrm O\eta'} &= 
\sqrt{\frac 1 5} 
\begin{pmatrix}
\hat 0 & \hat O^{\eta'} \hat E \\
\hat 0 & \hat 0
\end{pmatrix}
\label{eq:3_of_4}
, \\
\check O^{\bar {\mathrm O}\eta'} &= 
\sqrt{\frac 1 5} 
\begin{pmatrix}
\hat 0 & \hat 0 \\
\hat E^\dg \hat O^{\eta'\dg} & \hat 0
\end{pmatrix}
, 
\label{eq:4_of_4}
\end{align}
each of which corresponds to even-parity ($\xi'={\rm D}$) / odd-parity ($\xi'=\bar {\rm D}$) multipoles,
electron-pair ($\xi'={\rm O}$) / hole-pair ($\xi'=\bar {\rm O}$) amplitudes of original electrons.
Here the information of parity enters to the expression since the left-top block of $\check O$ originates from $\vec c_{\bm k}$ and right-bottom from $\vec c_{-\bm k}$.
Note that the further higher-order multipoles are not necessary.
These matrices
satisfy
the orthonormal relation
\begin{align}
\trace \left( \check O^{\xi_1'\eta_1'\dg} \check O^{\xi_2'\eta_2'} \right)
= \delta_{\xi_1'\xi_2'} \delta_{\eta_1'\eta_2'}
.
\end{align}
Then any $8\times 8$ matrix $\check A$ can be expanded as
\begin{align}
\check A &= \sum_{\xi'\eta'} a^{\xi'\eta'} \check O^{\xi'\eta'}
\end{align}
and the expansion coefficients are extracted as
\begin{align}
a^{\xi'\eta'} &=  \trace \left( \check O^{\xi'\eta'\dg} \check A \right).
\end{align}
Hence the series of the above matrices (\ref{eq:1_of_4}--\ref{eq:4_of_4}) can be regarded as complete.
With this property one can show the relation
\begin{align}
\sum_{\xi'\eta'} \trace (\check O^{\xi'\eta'\dg} \check A)\, \trace (\check O^{\xi'\eta'} \check B)
= \trace (\check A \check B)
\end{align}
for arbitrary matrices $\check A$ and $\check B$.
This relation is useful in confirming the sum rule \eqref{eq:sum_orig} of the main text.

\section{$\bm k$-dependent form factors}
In Eqs.~\eqref{eq:def_diag} and \eqref{eq:def_pair}, we consider $\bm k$-dependent form factor
\begin{align}
g_\eta(\bm k) = g'_\eta(k,k_z) f_\eta(\phi)
\end{align} 
with the cylindrical coordinate system $\bm k=(k,\phi,k_z)$.
The complete functional forms are given by
\begin{align}
g'_{1} &= \sqrt{\frac{1}{4\pi}}
,\ f_{1} = 1
\end{align}
for $s$-wave,
\begin{align}
g'_{y} &= \sqrt{\frac{3}{4\pi}}\, \frac{k}{|\bm k|}
,\ f_{y} = \sin \phi
\\
g'_{z} &= \sqrt{\frac{3}{4\pi}}\, \frac{k_z}{|\bm k|}
,\ f_{z} = 1
\\
g'_{x} &= \sqrt{\frac{3}{4\pi}}\, \frac{k}{|\bm k|}
,\ f_{x} = \cos \phi
\end{align}
for $p$-wave,
\begin{align}
g'_{xy} &= \sqrt{\frac{15}{16\pi}}\, \frac{k^2}{|\bm k|^2}
,\ f_{xy} = \sin 2\phi
\\
g'_{yz} &= \sqrt{\frac{15}{4\pi}}\, \frac{kk_z}{|\bm k|^2}
,\ f_{yz} = \sin \phi
\\
g'_{3z^2-r^2} &= \sqrt{\frac{5}{16\pi}}\, \frac{3k_z^2-|\bm k|^2}{|\bm k|^2}
,\ f_{3z^2-r^2} = 1
\\
g'_{zx} &= \sqrt{\frac{15}{4\pi}}\, \frac{kk_z}{|\bm k|^2}
,\ f_{zx} = \cos \phi
\\
g'_{x^2-y^2} &= \sqrt{\frac{15}{16\pi}}\, \frac{k^2}{|\bm k|^2}
,\ f_{x^2-y^2} = \cos 2\phi
\end{align}
for $d$-wave, and
\begin{align}
g'_{y(3x^2-y^2)} &= \sqrt{\frac{35}{32\pi}}\, \frac{k^3}{|\bm k|^3}
,\ f_{y(3x^2-y^2)} = \sin 3 \phi
\\
g'_{xyz} &= \sqrt{\frac{105}{16\pi}}\, \frac{k^2k_z}{|\bm k|^3}
,\ f_{xyz} = \sin 2 \phi
\\
g'_{y(5z^2-r^2)} &= \sqrt{\frac{21}{32\pi}}\, \frac{k(5k_z^2-|\bm k|^2)}{|\bm k|^3}
,\ f_{y(5z^2-r^2)} = \sin \phi
\\
g'_{z(5z^2-3r^2)} &= \sqrt{\frac{7}{16\pi}}\, \frac{k_z(5k_z^2-3|\bm k|^2)}{|\bm k|^3}
,\ f_{z(5z^2-3r^2)} = 1
\\
g'_{x(5z^2-r^2)} &= \sqrt{\frac{21}{32\pi}}\, \frac{k(5k_z^2-|\bm k|^2)}{|\bm k|^3}
,\ f_{x(5z^2-r^2)} = \cos \phi
\\
g'_{z(x^2-y^2)} &= \sqrt{\frac{105}{16\pi}}\, \frac{k^2k_z}{|\bm k|^3}
,\ f_{z(x^2-y^2)} = \cos 2\phi
\\
g'_{x(x^2-3y^2)} &= \sqrt{\frac{35}{32\pi}}\, \frac{k^3}{|\bm k|^3}
,\ f_{x(x^2-3y^2)} = \cos 3 \phi
\end{align}
for $f$-wave,
where $|\bm k| = \sqrt{k^2+k_z^2}$.
The coefficients are determined by the normalization condition 
\begin{align}
\int \diff \Omega_{\bm k} |g_\eta(\bm k)|^2 = 1
,
\end{align}
where $\int \diff \Omega_{\bm k}$ means the integral over the spherical surface.
Note that one can consider the further higher-order functions since there are infinite number of degrees of freedom in $\bm k$-space.
If one uses the Fourier expansion by utilizing the function $f_\eta(\phi)$, one can in principle decompose $\chi_{\xi\eta}^{\xi'\eta'}$ further depending on the rotational symmetry breaking in $xy$-plane.

\end{document}